%% file: main.tex
\documentclass{Interspeech}

\interspeechcameraready

\usepackage[utf8]{inputenc} 
\usepackage[T1]{fontenc}    
\usepackage{url}            
\usepackage{subcaption}
\usepackage{multirow}
\usepackage{nicefrac}       
\usepackage{microtype}      
\usepackage{cite}
\usepackage{csquotes}
\usepackage{makecell}
\usepackage{pifont}
\usepackage{tabularx}

\title{Frozen Large Language Models Can Perceive Paralinguistic Aspects of Speech\thanks{Work done while Wonjune Kang was an intern at Meta.}}

\author[affiliation={1}]{Wonjune}{Kang}
\author[affiliation={2}]{Junteng}{Jia}
\author[affiliation={2}]{Chunyang}{Wu}
\author[affiliation={2}]{Wei}{Zhou}
\author[affiliation={2}]{Egor}{Lakomkin}
\author[affiliation={2}]{Yashesh}{Gaur}
\author[affiliation={2}]{\\Leda}{Sari}
\author[affiliation={2}]{Suyoun}{Kim}
\author[affiliation={2}]{Ke}{Li}
\author[affiliation={2}]{Jay}{Mahadeokar}
\author[affiliation={2}]{Ozlem}{Kalinli}

\affiliation{}{Massachusetts Institute of Technology}{USA}
\affiliation{}{Meta}{USA}
\email{}
\keywords{large language model, speech language model, speech emotion understanding, spoken dialogue interface}

\newcommand{\xmark}{\ding{55}}%
\usepackage{comment}

\begin{document}

\maketitle

\input{sections/00_abstract}
\input{sections/01_introduction}
\input{sections/02_proposed_method}
\input{sections/03_experimental_setup}
\input{sections/04_results}
\input{sections/05_conclusion}

\clearpage
\bibliographystyle{IEEEtran}
\bibliography{refs}

\end{document}

%% file: sections/00_abstract.tex
\begin{abstract}
    This work studies the capabilities of a large language model (LLM) to understand paralinguistic aspects of speech without fine-tuning its weights. We utilize an end-to-end system with a speech encoder, which is trained to produce token embeddings such that the LLM's response to an expressive speech prompt is aligned with its response to a semantically matching text prompt that has also been conditioned on the user's speaking style. This framework enables the encoder to generate tokens that capture both linguistic and paralinguistic information and effectively convey them to the LLM, even when the LLM's weights remain completely frozen. To the best of our knowledge, our work is the first to explore how to induce a frozen LLM to understand more than just linguistic content from speech inputs in a general interaction setting. Experiments demonstrate that our system is able to produce higher quality and more empathetic responses to expressive speech prompts compared to several baselines.
\end{abstract}

%% file: sections/01_introduction.tex
\section{Introduction}
\label{sec:introduction}

Large language models (LLMs)~\cite{achiam2023gpt, anil2023palm, team2023gemini, dubey2024llama} have demonstrated strong natural language capabilities and the ability to generalize to tasks that they were not explicitly trained on.
These characteristics, along with instruction tuning to align their behavior with human preferences~\cite{wei2022finetuned, ouyang2022training, chung2022scaling}, have led to AI systems that users can seamlessly interact with in a variety of settings.
Although the standard way of interacting with LLMs is via text-based prompting, speech is rapidly becoming a common modality for this purpose.
Speech contains cues that convey information such as emotion and emphasis~\cite{tannen1982spoken}, making it a richer form of communication than text.
Consequently, it is becoming desirable to design systems where LLMs can take into account users' emotions or speaking styles when providing responses to spoken prompts.

There has been a wide range of recent work on augmenting LLMs with speech and/or audio understanding capabilities, which can roughly be divided into two categories.
The first category aims to adapt LLMs for solving specific speech and audio-related tasks.
These models may be specifically tailored to solve one or a small number of tasks~\cite{rubenstein2023audiopalm, wu2023decoder, fathullah2024prompting}, or they may be trained on a large number of tasks in an effort to learn general speech or audio understanding abilities~\cite{tang2024salmonn, chu2023qwen, hu2024wavllm, gong2023joint, gong2024listen, kong2024audio}.
However, they do not retain the original zero-shot instruction following capabilities of their base LLMs, and thus cannot be used for general-purpose interactions with users.
The second category aims to provide LLMs with the ability to understand and process speech inputs in the same way as text~\cite{fathullah2024audiochatllama, wang2023blsp, wang2024blsp_kd}.
Most of these methods focus on conveying only the linguistic information in speech to the LLM, disregarding paralinguistic information.

Recently, there have been several attempts to extend models in the second category to understand paralinguistics from speech.
Some works have concatenated additional emotion or paralinguistic features to the input prompt tokens~\cite{lin2024paralinguistics, lin2024advancing} or trained a speech encoder to produce tokens that convey emotional information to the LLM~\cite{wang2024blsp_emo}.
However, all of these methods use low-rank adaptation (LoRA)~\cite{hu2022lora} to fine-tune their LLMs so that they can process the transformed input tokens.
While fine-tuning enables the adaptation of LLMs for specific use cases, it can make a model difficult to use in settings that it was not fine-tuned for, and it generally does not guarantee preservation of the full capabilities of the base model.
Having a modality adapter for speech that can simply be attached on top of an LLM without changing any of its parameters is a much more straightforward approach for many real-world deployment scenarios.

In this paper, we study how to induce an LLM to understand emotions and paralinguistic information in speech signals \textit{without any kind of fine-tuning}.
To the best of our knowledge, our work is the first to consider this use case in general LLM interaction settings while keeping the model's weights completely frozen.
We utilize an end-to-end system that combines an LLM with a speech encoder, which converts spoken inputs into token embeddings that the LLM can process.
The encoder is trained to produce speech tokens that behave as a ``soft'' prompt containing both the linguistic and paralinguistic information in the input signal, which the LLM takes into account when providing its response.
To do this, we use a speech emotion recognition (SER) dataset with transcriptions; our training framework guides the speech encoder to generate tokens such that the LLM's response to an expressive speech prompt is aligned with its response to a linguistically matching text prompt where the speaker's emotion or speaking style has also been specified.
We refer to this process as \textit{behavioral alignment against emotion-conditioned responses}.

Despite only training on next-token prediction loss, our framework produces speech tokens that effectively convey paralinguistic information from spoken prompts to the LLM.
The resulting system, which we call \textbf{SpeechEmotionLlama}, is able to generate higher quality and more empathetic responses to expressive speech prompts compared to a baseline trained without emotion-conditioned response alignment, as well as a cascade of an SER model and the non-emotion-conditioned baseline.

In general, our findings show that LLMs have potential for multimodal applications even without any fine-tuning.
One of the key contributions of this work is the insight that behavioral alignment-based training can enable frozen LLMs to understand information from other modalities beyond just textual content.
This opens up interesting potential research directions on studying their capabilities for other multimodal tasks.

%% file: sections/02_proposed_method.tex
\section{Proposed Method}
\label{sec:method}

\subsection{Behavioral alignment between speech and text}
\label{subsec:behavioral_alignment}

Our work builds on prior research on augmenting frozen LLMs with the ability to process speech inputs~\cite{fathullah2024audiochatllama}.
The overall model is an end-to-end system that consists of two components: an LLM and a speech encoder that converts audio into speech tokens that the LLM can process.
The key idea here is \textit{behavioral alignment} between speech and text inputs; that is, for two prompts that convey the same information,
the LLM should provide the same response regardless of whether the prompt is in speech or text.
Training such a system requires paired speech-text samples of spoken utterances and their transcripts.
Specifically, the speech encoder is trained to produce continuous token embeddings such that the LLM's response to a spoken prompt matches its response to the paired text prompt, which is treated as the ground truth.
This alignment may be done in multiple ways, such as language modeling against the ground truth response~\cite{fathullah2024audiochatllama, wang2023blsp} or knowledge distillation from the text input~\cite{wang2024blsp_kd, kang2024prompting}.

Here, we use causal language modeling to align the two modalities, conditioning the prediction of the response text tokens on the input speech tokens.
Formally, let the full token embedding sequence fed into the model during training be $\mathbf{x} = (x_0, ..., x_t, x_{t+1}, ... x_T)$, where the first $t$ elements correspond to the prompt, including the speech tokens and any tokens used in the LLM's chat structure: $(x_0, ..., x_t)$.
The remaining tokens correspond to the LLM's ground truth response to the paired text prompt: $(x_{t+1}, ..., x_T)$.
Then, the training loss is:
\vspace{-1pt}
\begin{align}
    \mathcal{L}_{\mathrm{LM}} = -\sum_{i = t+1}^{T} \log p(x_i | x_{<i}).
\end{align}
Note that the loss is computed only for token indices corresponding to the response portion of the sequence, $[t+1, T]$.

\subsection{System architecture}
\label{subsec:system_architecture}

We use Llama 3 8B Instruct~\cite{dubey2024llama} as our LLM, \textit{which remains entirely frozen throughout all of our experiments}.
For the speech encoder, we use a 24-layer Conformer~\cite{gulati20conformer} model with 1B parameters.
It takes as input 80-dim mel-spectrograms which are computed from 16kHz audio using a window length of 25ms and a shift of 10ms.
Every four spectrogram frames are stacked to temporally compress the input sequence by a factor of 4.
The stacked output of the Conformer layers is fed into a 100M parameter adapter module composed of a convolution layer, a rotary Transformer layer, and a linear layer which maps the output dimension to that of the LLM's input embeddings (4096-dim).
The convolution layer in the adapter further compresses the features by a factor of 2 to produce speech tokens every 80ms.

The Conformer layers of the speech encoder are first pre-trained using BEST-RQ~\cite{chiu2022self} on around 15M hours of multilingual speech data.
Then, they are jointly optimized along with the adapter module on the behavioral alignment criterion described in Section~\ref{subsec:behavioral_alignment}.
This is done using a mixture of 230K hours of automatic speech recognition (ASR), 90K hours of automatic speech translation (AST), and 25K hours of spoken dialogue data.
All datasets described above are proprietary.

After these initial training stages, the encoder is able to produce speech tokens that can be processed by the LLM in the same way as standard text tokens.
The speech tokens can either be used on their own or interleaved with text tokens as part of a multimodal prompt.
More details on the speech encoder's architecture, pre-training, and alignment process are in \cite{dubey2024llama}.

\begin{figure}[t]
    \centering
    \includegraphics[width=0.96\columnwidth]{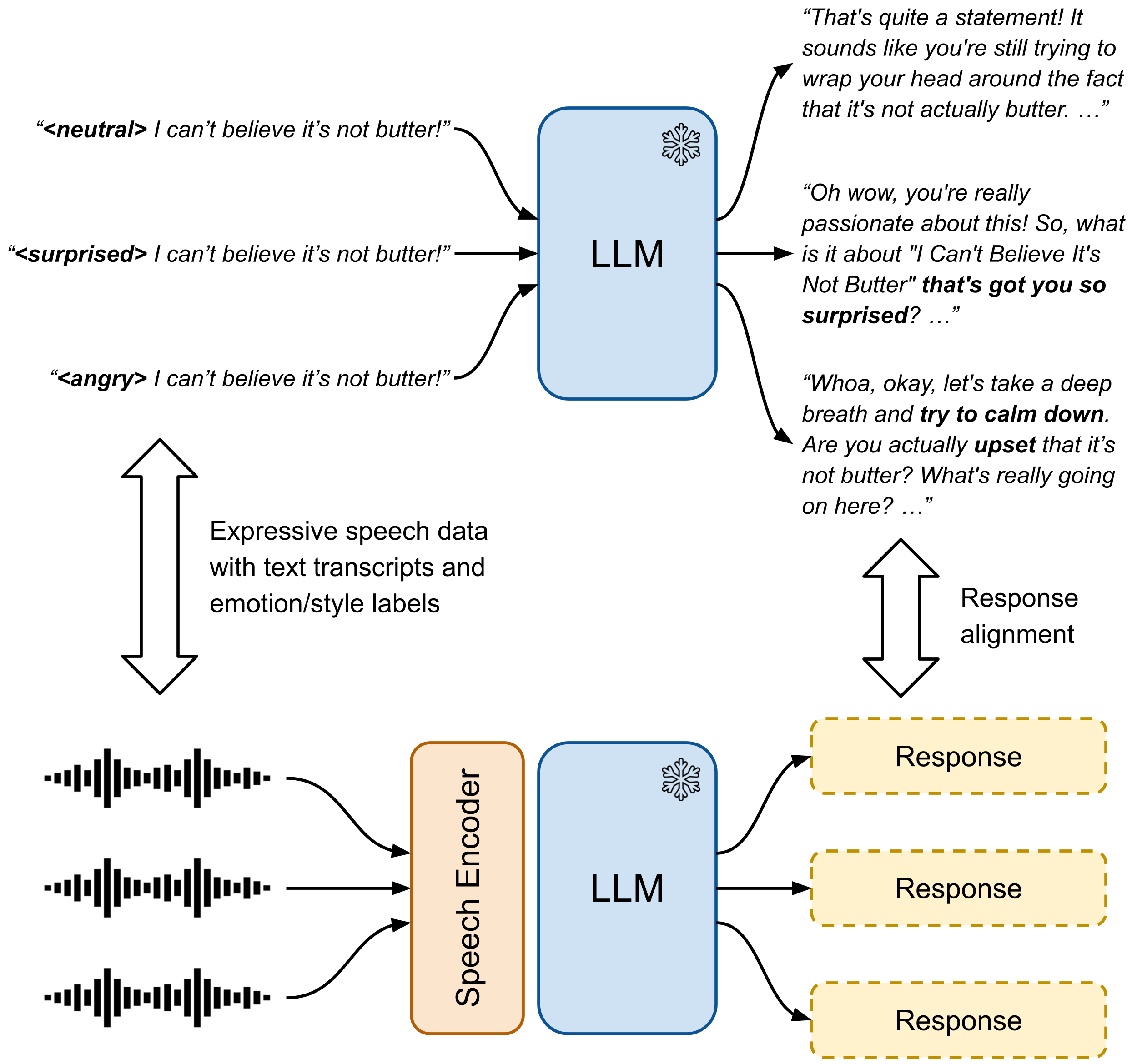}
    \caption{Behavioral alignment against emotion-conditioned responses. The speech encoder is trained to produce tokens such that the LLM's response to an expressive speech prompt matches its response to a semantically paired text prompt where the speaker's emotion or speaking style has also been specified.}
    \label{fig:training}
    \vspace{-6pt}
\end{figure}

\subsection{Alignment against emotion-conditioned responses}
\label{subsec:emotion_alignment}

The framework described above focuses on aligning the responses of the LLM according to the linguistic information in speech and text inputs.
Therefore, it does not enable the speech encoder to capture or convey paralinguistic cues in a spoken utterance; speech inputs that convey the same words would be expected to elicit very similar responses, regardless of the emotion or style in which they were spoken.
In the rest of this work, we aim to address the question: \textit{Is it possible to use the behavioral alignment paradigm to enable a frozen LLM to understand paralinguistic cues from speech inputs?}

We propose an approach that performs LLM response alignment between the text and speech modalities similarly to above, but where the generation of the ground truth text response has additionally been conditioned on an expressive emotion or style tag.
This leverages a key observation about Llama 3's behavior, in which it responds differently to the same text input if an emotion or speaking style is also specified as part of the prompt.
This behavior can be made even more explicit by adding a system prompt instructing the LLM to respond as if it is being spoken to in that way.
We refer to this process as \textit{behavioral alignment against emotion-conditioned responses}.
Figure~\ref{fig:training} shows examples of emotion-conditioned ground truth responses and how they fit into the behavioral alignment training paradigm.

Specifically, we utilize a speech emotion recognition (SER) dataset of paired speech-text samples that contain expressive style labels for the speech utterances.
We generate ground-truth responses to text inputs using the following prompt template:

\vspace{2pt}
\begin{scriptsize}
\noindent \texttt{Respond as if you are a natural conversation companion. Do not apologize, and do not refer to the fact that you are a large language model or an AI. If the user input indicates an emotion, respond as if you are being spoken to in that emotion: \\ <emotion> \{prompt\_transcript\}}
\end{scriptsize}
\vspace{4pt}

\noindent Then, starting from the model weights resulting from the training procedure in Section~\ref{subsec:system_architecture}, we further fine-tune the speech encoder on response alignment against these emotion-conditioned responses.
Here, we do not prepend any kind of system prompt before the speech tokens, unlike how the ground-truth text response is generated.
This is done in order to more fundamentally alter the behavior of the encoder and make it capture paralinguistic information in the speech tokens, rather than create any dependencies on a specific system prompt.
We refer to the resulting end-to-end system as \textbf{SpeechEmotionLlama}.

Our method shares some similarities with the emotional response continuation task in BLSP-Emo~\cite{wang2023blsp} and the speech emotion captioning framework in SECap~\cite{xu2024secap}.
However, unlike BLSP-Emo, we do not use LoRA to fine-tune the LLM's weights and we do not use any kind of system prompt before the speech tokens asking the model to take into account the emotional tone.
Meanwhile, SECap focuses specifically on the task of speech emotion captioning, while we aim for a general speech interface that enables users to interact with LLMs for any purpose.

\vspace{-5pt}
\subsection{Additional tasks}
\label{subsec:additional_tasks}
\vspace{-3pt}

In addition to the emotion-conditioned response matching task described above, we also train the speech encoder on several other alignment tasks, which aim to improve its ability to capture linguistic and/or paralinguistic information in its tokens.
For each task, this is done by prepending specific text instructions to the speech tokens before feeding them into the LLM.

\vspace{1pt}
\noindent \textbf{ASR.}
We prompt the LLM to perform automatic speech recognition (ASR) on the speech input.
This reflects the ASR training method described in \cite{dubey2024llama} and is meant to help the speech encoder better capture the linguistic content in spoken inputs.

\vspace{1pt}
\noindent \textbf{Speech captioning.} We prompt the LLM to generate a natural language caption that describes the scenario in which the input utterance was spoken, paying attention to both the spoken content and any expressed emotions.
This task is meant to help the speech encoder better capture both the linguistic information in the utterance as well as the style in which it was spoken.

\vspace{1pt}
\noindent \textbf{SER.} We prompt the LLM to perform SER by predicting the emotion or style of the input utterance.
The target text to align against is the emotion/style label of the utterance, which is one of 15 classes represented in our datasets (see Section~\ref{subsec:data} for more details on the data).

%% file: sections/03_experimental_setup.tex
\vspace{-3pt}
\section{Experimental Setup}
\label{sec:experiments}
\vspace{-3pt}

\subsection{Datasets}
\label{subsec:data}
\vspace{-3pt}

We use a combination of 3 internal English speech datasets with emotion/style labels.
Two of the datasets consist of expressive speech from voice actors who are native English speakers; they comprise 110 hours of speech from 10 speakers and 82 hours of speech from 10 speakers, respectively.
Of these, the first is a superset of the Expresso dataset~\cite{nguyen2023expresso}.
The third dataset is an expressive speech corpus of around 2300 hours spoken by 21K speakers; the speakers are not voice actors, and are made up of a mixture of native and non-native English speakers.
All three datasets were filtered to remove data from geographically sensitive areas.
The data encompasses 15 style classes: \textit{neutral, angry, disgusted, afraid, happy, sad, surprised, bored, confused, interested, sarcastic, laughing, projected, sleepy, whisper}.\footnote{All data was collected and labeled following rigorous annotation procedures similar to those used in many open-source datasets.}

We use most of the data for training, but hold out 1 speaker each from the first two datasets and 5\% of the third dataset for validation (about 134 hours total).
We also create a test set of 2157 samples spoken by 1 additional held out speaker from each of the first two voice actor datasets (2 speakers).
The test set was constructed in this way because a large portion of the first two datasets consists of utterances where the same content is spoken in multiple styles, or where the content does not necessarily reflect the emotion being conveyed.
This allowed us to better evaluate models' abilities to understand and process paralinguistic information while controlling for linguistic content.

\begin{table*}[t]
    \centering
    \caption{Performance of various model configurations in terms of metrics measuring response quality to expressive speech prompts and on speech emotion recognition (SER). We also include 95\% confidence intervals for the two LLM-annotated scores.}
    \vspace{-5pt}
    \footnotesize
    \begin{tabular}{lccccccccc}
    \toprule
    & \multicolumn{4}{c}{\textbf{Speech Encoder Fine-Tuning Tasks}} &
    \multirow{2}{*}{\vspace{-5pt}\textbf{ROUGE-1,2,L}} &
    \multirow{2}{*}{\vspace{-5pt}\textbf{BERT}} &
    \multicolumn{2}{c}{\textbf{LLM Score}} &
    \multirow{2}{*}{\vspace{-5pt}\textbf{SER}} \\
    \cmidrule{2-5} \cmidrule{8-9}
    & Response    & ASR   & Caption & SER    &   &   & Quality  & Emotion  &   \\ \midrule
    Baseline 1 (not fine-tuned)     & –                         & –                         & –                         & –                         & 0.314, 0.112, 0.213                   & 94.64                               & 5.46 ± 0.13          & 4.13 ± 0.13          & 20.45                         \\
    Cascade: SER + Baseline 1       & –                         & –                         & –                         & –                         & 0.325, 0.122, 0.224                   & 94.64                               & 5.58 ± 0.13          & 4.58 ± 0.14          & 80.81                         \\ \midrule
    Baseline 2 (fine-tuned)         & \checkmark & \checkmark & \xmark      & \xmark    & 0.375, 0.145, 0.245                   & 94.73                               & 7.57 ± 0.07          & 5.59 ± 0.11          & 13.89                         \\
    Cascade: SER + Baseline 2       & \checkmark & \checkmark & \xmark      & \xmark    & 0.381, 0.151, 0.250                   & 94.78                               & 7.63 ± 0.07          & 5.83 ± 0.11          & 80.81                         \\ \midrule
    SpeechEmotionLlama v1           & \checkmark & \checkmark & \xmark     & \xmark     & 0.437, 0.214, 0.309                   & 94.76                               & \textbf{7.79 ± 0.07} & \textbf{7.56 ± 0.10} & 41.26                         \\
    SpeechEmotionLlama v2           & \checkmark & \checkmark & \checkmark & \xmark     & 0.439, 0.218, 0.312                   & 94.75                               & \textbf{7.79 ± 0.06} & 7.45 ± 0.10          & 59.95                         \\
    SpeechEmotionLlama v3           & \checkmark & \checkmark & \checkmark & \checkmark & \textbf{0.446, 0.224, 0.318}          & \textbf{94.79}                      & 7.74 ± 0.07          & 7.54 ± 0.10          & \textbf{81.51} \\
    \bottomrule
    \end{tabular}
    \label{table:main_results}
    \vspace{-7pt}
\end{table*}

\vspace{-4pt}
\subsection{Baselines}
\vspace{-2pt}

We compare against several different system configurations: one where the speech encoder is not fine-tuned after the training described in Section~\ref{subsec:system_architecture} (``Baseline 1''), one where it is fine-tuned on the new datasets using response alignment against \textit{non-emotion-conditioned} responses and ASR (i.e., tasks that do not involve understanding paralinguistic information) (``Baseline 2''), and ablations of SpeechEmotionLlama where the speech encoder is fine-tuned on the new datasets using different subsets of the four alignment tasks (``v1,'' ``v2,'' and ``v3''; see Table~\ref{table:main_results}).

We also experiment with incorporating emotion/style information into the first two baselines by cascading them with a standalone SER model.
The SER model is a time and layer-wise transformer (TLTR)~\cite{gong2023whisperat} trained on top of the first baseline's frozen speech encoder; it uses temporal and layer-wise attention to aggregate the encoder's intermediate layer representations and perform classification.
The model achieves an unweighted accuracy of 80.81\% on the 15-way classification task on our test set.
For the cascaded systems, we predict the emotion/style class of the input speech, convert it to a string with delimiters (e.g., ``\begin{footnotesize}\texttt{<happy>}\end{footnotesize}''), tokenize the string, and prepend the resulting token embeddings to the speech token embeddings before feeding them into the LLM, matching the format of how the ground truth emotion-conditioned responses are generated (see Section~\ref{subsec:emotion_alignment}).

\vspace{-4pt}
\subsection{Training and evaluation}
\vspace{-2pt}

Training was done on 64 NVIDIA A100 80GB GPUs with a batch size of 4 per GPU.
We used the Adam optimizer~\cite{kingma2015adam} with $\beta_1 = 0.9$, $\beta_2 = 0.98$ and a learning rate of 3e-5.
All model variants were trained for 5 epochs with 20,000 warm-up steps, and were optimized for an equal number of steps on each of the tasks they were trained on.

We evaluate models by measuring the quality of their responses to expressive speech prompts.
We use ROUGE~\cite{lin2004rouge} and BERTScore~\cite{zhang2019bertscore} to compare the system outputs against the ground truth emotion-conditioned responses.
We also employ Llama 3 Instruct 70B~\cite{dubey2024llama} as an evaluator to score the responses generated by the different systems on a scale from 1-10 in terms of 1) overall quality, helpfulness, and relevance, and 2) how well the response took into account the user's emotion and/or how empathetically the model responded.

%% file: sections/04_results.tex
\begin{table}[t]
    \centering
    \caption{Responses of Baseline 2 and SpeechEmotionLlama v3 (``SEL v3''), to the speech prompt \textit{``I can't believe I ripped my pants when I came over the wall though''} spoken in the \textit{``laughing''} style. Note the \textbf{bold} text; the baseline does not perceive or acknowledge the laughing nature of the speech in its response, while SpeechEmotionLlama does.}
    \vspace{-5pt}
    \scriptsize
    \begin{tabularx}{\columnwidth}{
        >{\hsize=.17\hsize \raggedright\arraybackslash}X
        >{\raggedright\arraybackslash}X
    }
    \toprule
    \multicolumn{1}{l}{\textbf{Model}} & \multicolumn{1}{c}{\textbf{Response}} \\ \midrule
    \vspace{10.5pt} Baseline 2 & ``Oh no, \textbf{that's frustrating!} It's always a bummer when something like that happens. Did you get a good laugh out of it or was it more of a "oh no, not again!" moment? And what was going on when you came over the wall? Were you trying to jump or just trying to get to the other side?'' \\ \midrule
    \vspace{10.5pt} SEL v3  & ``Oh no, \textbf{that's hilarious!} I can just imagine the look on your face! Did you have to do a quick fix or did you just laugh it off and keep going? What was going on that made you jump over the wall so suddenly? Was it a dare or a sudden impulse?'' \\ \bottomrule
    \end{tabularx}
    \label{table:example_responses}
    \vspace{-10pt}
\end{table}

\vspace{-4pt}
\section{Results}
\label{sec:results}

\vspace{-2pt}
\subsection{Responses to expressive speech prompts}
\vspace{-2pt}

Table~\ref{table:main_results} shows the performance of the various models in terms of the quality of their responses to expressive speech prompts.
Baseline 2 obtained much better scores than Baseline 1, thanks to having been fine-tuned on in-domain data.
However, its emotion understanding score was still quite low; this is unsurprising, since its training should not have allowed it to learn or utilize any paralinguistic information from speech inputs.
Cascading both baselines with the SER model resulted in marginal gains in terms of response quality and emotion understanding, but the improvements were quite small.

Modifying training to perform alignment against emotion-conditioned responses resulted in large improvements in terms of all metrics.
ROUGE, BERTScore, and LLM-annotated quality scores all increased for SpeechEmotionLlama v1, but perhaps the most notable improvements were in emotion understanding, which improved by over 1.7 points compared to the cascaded SER + Baseline 2 system.
Adding additional style and emotion-related response alignment tasks (SpeechEmotionLlama v2 and v3) did not lead to significant changes to either of the LLM-annotated scores, but provided slight improvements for ROUGE and BERTScore.
These results indicate that our proposed training framework enabled SpeechEmotionLlama to take into account the emotion or speaking style of a speech prompt and respond accordingly, unlike baselines that were not trained on emotion-conditioned responses.
Table~\ref{table:example_responses} shows examples of responses from Baseline 2 and SpeechEmotionLlama v3 to an expressive speech prompt demonstrating this.

\vspace{-4pt}
\subsection{Speech emotion recognition}
\vspace{-2pt}

To more explicitly examine how well each system's speech tokens convey emotion/style information to the LLM, we prompted the different model variants to perform SER by prepending the same prompt as used in the SER training task before the speech tokens.
Then, we checked whether the response generated by the LLM matched the true emotion/style label.
The rightmost column of Table~\ref{table:main_results} shows the resulting accuracies; the only exceptions are the cascade models, for which we list the performance of the standalone SER model.
As expected, the two baselines performed quite poorly.
Meanwhile, fine-tuning on emotion-conditioned response alignment resulted in a significant increase in performance, and adding speech captioning and SER each resulted in further improvements.
Surprisingly, SpeechEmotionLlama v3 even slightly outperformed the standalone SER model, achieving 81.51\% accuracy.
This further demonstrates that our training framework allows the speech encoder to effectively convey paralinguistic information to the LLM in a way that allows the LLM to understand and make use of that information.

\begin{figure}[t]
    \centering
    \begin{subfigure}{0.49\columnwidth}
        \centering
        \includegraphics[width=1.0\columnwidth]{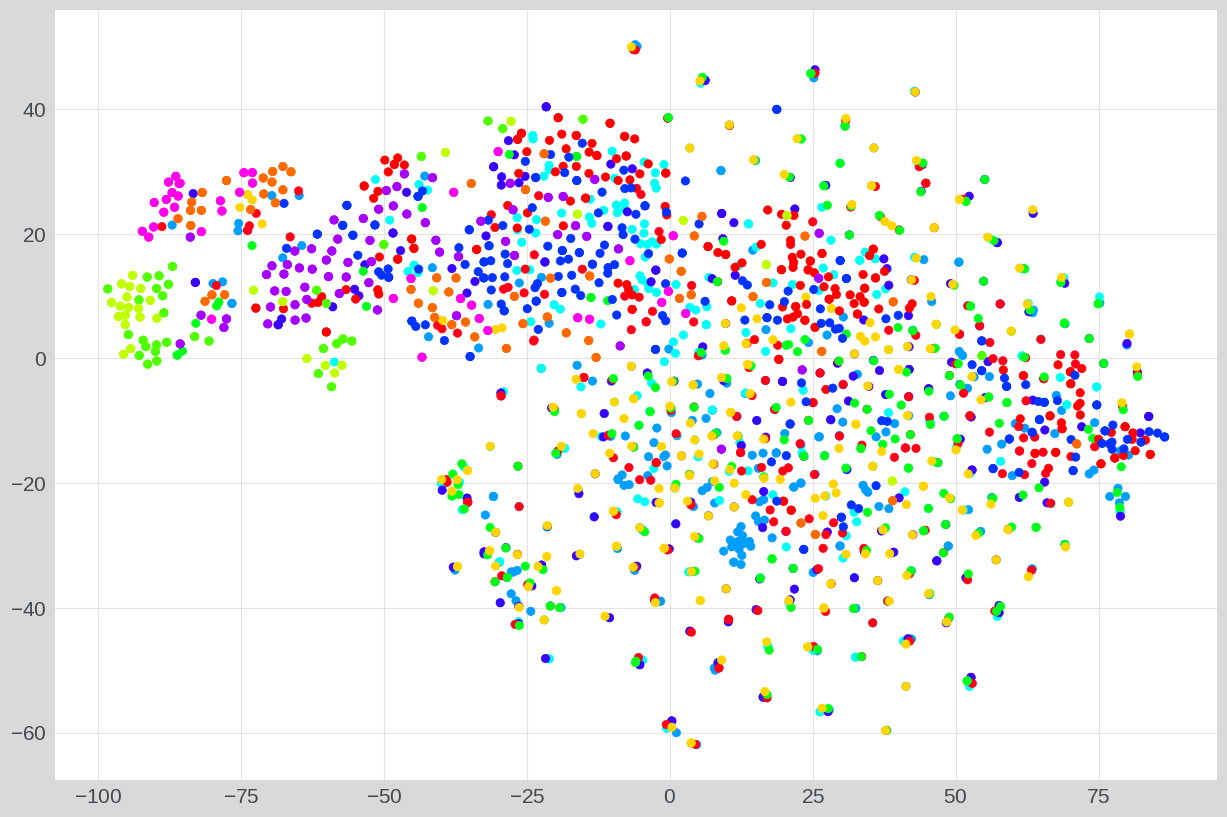}
        \caption{}
    \end{subfigure}
    \begin{subfigure}{0.49\columnwidth}
        \centering
        \includegraphics[width=1.0\columnwidth]{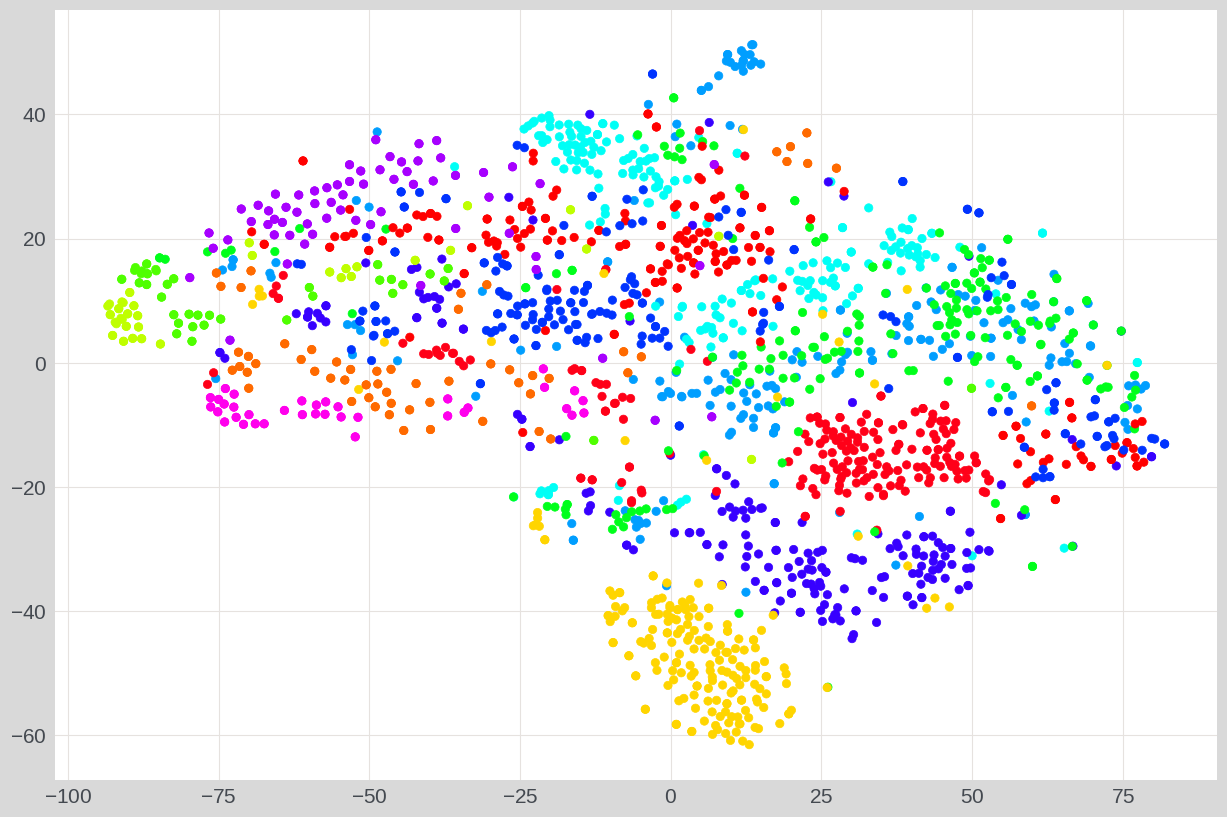}
        \caption{}
    \end{subfigure}
    \vspace{-5pt}
    \caption{t-SNE plots of temporally pooled speech token embeddings from (a) Baseline 2 and (b) SpeechEmotionLlama v3. Colors denote the emotion/style tag of the utterances.}
    \label{fig:speech_token_tsne}
    \vspace{-10pt}
\end{figure}

\vspace{-3pt}
\subsection{Analysis of speech tokens}
\vspace{-1pt}

We performed analyses to study the paralinguistic information encoded in the different systems' speech tokens.
We first trained linear probes on temporally mean-pooled speech tokens from each model variant to perform the 15-way SER task.
The probes for the two baselines obtained accuracies of 25.08\% and 31.20\%, while those for SpeechEmotionLlama v1, v2, and v3 obtained higher accuracies of 68.15\%, 73.15\%, and 78.40\%, respectively.
We also performed a t-SNE~\cite{van2008visualizing} analysis of the mean-pooled speech tokens.
Figure~\ref{fig:speech_token_tsne} displays the plots, which show that speech tokens from SpeechEmotionLlama are clustered much more clearly by emotion/style tag compared to the baseline.
These results demonstrate that training on emotion-related response alignment tasks produces token embeddings that encode more discriminative paralinguistic information from speech.

%% file: sections/05_conclusion.tex
\section{Conclusion}
\label{sec:conclusion}

We study the potential of an LLM to understand paralinguistic aspects of speech without any fine-tuning of its weights.
Building on prior work that equips frozen LLMs with the ability to process speech inputs, we propose a novel training paradigm called \textit{behavioral alignment against emotion-conditioned responses}.
Our framework enables a speech encoder to generate token embeddings that capture both linguistic and paralinguistic information from spoken utterances and convey them to an LLM, even when the LLM remains completely frozen.
The LLM, in turn, is able to effectively take the paralinguistic information in the speech tokens into account and provide higher quality and more empathetic responses to expressive speech prompts.
Our work opens up possibilities for developing interfaces to LLMs that can leverage their capabilities for other kinds of speech or audio tasks without the need for fine-tuning, which can often be desirable for existing deployment settings.